\documentclass[pra,showpacs,amsmath,amssymb,superscriptaddress]{revtex4}
\usepackage{graphicx}
\usepackage{bm}
\usepackage{amsmath,amssymb}

\begin{document}

\title{
Protected Rabi oscillation induced 
by natural interactions among physical qubits
}

\author{Naoaki Kokubun}
\email{kokubun@ASone.c.u-tokyo.ac.jp}
\affiliation{
Department of Basic Science, University of Tokyo, 
3-8-1 Komaba, Tokyo 153-8902, Japan
}
\author{Akira Shimizu}
\email{shmz@ASone.c.u-tokyo.ac.jp}
\affiliation{
Department of Basic Science, University of Tokyo, 
3-8-1 Komaba, Tokyo 153-8902, Japan
}
\date{\today}
\begin{abstract}
For a system composed of nine qubits, we show that natural 
interactions among the qubits induce the time evolution that 
can be regarded, at discrete times, as the Rabi oscillation 
of a logical qubit.  Neither fine tuning of the parameters 
nor switching of the interactions is necessary.  Although 
straightforward application of quantum error correction fails, 
we propose a protocol by which the logical Rabi oscillation is 
protected against all single-qubit errors. 
The present method thus opens 
a simple and realistic way of protecting the unitary time 
evolution against noise.
\end{abstract}
\maketitle

\section{Introduction}

Decoherence of quantum states has been attracting much attention 
for long years \cite{SM02}.
Many methods have been proposed for defeating the decoherence.
As compared with other methods \cite{ZR97,KBLW01,VKL99,Facchi05},
quantum error correction (QEC) \cite{Sh95,St96,NC00,Go97,Pre98}
has a great advantage of 
protecting against arbitrary errors 
if they only affect a single 
qubit (two-level system) in each {\em logical} qubit \cite{NC00}.
Although QEC has been developed in the context of 
quantum computation, 
it is interesting and useful to apply QEC
to protection of the unitary time evolution 
(Hamiltonian evolution) against noise.

When trying to realize this,
however, 
one encounters many physical problems,
which are usually disregarded in 
discussions on the computational complexity \cite{NC00}.
For example, 
some physical process may be much more difficult to realize than another, 
even if the number of the necessary steps for them differs
`only by polynomial steps'  \cite{NC00}.
Furthermore, 
fabrication of a controlled-NOT gate, 
which is one of the elementary quantum gates, 
is very difficult because it requires 
fine tuning of the coupling constants 
of the interactions 
and high-precision switching of them, 
even if one employs the excellent ideas of Refs.~\cite{DiV00,ML05}.
Assembling a quantum circuit from the elementary gates 
is even more difficult, 
particularly when the circuit is large and complicated.
Unfortunately, the circuit indeed becomes large and complicated
when one tries to apply QEC to 
the Hamiltonian evolution, even for the simplest case
such as the Rabi oscillation \cite{Pre98}. 
The largest and most complicated part of the circuit is the one that 
induces the encoded Hamiltonian evolution 
(such as the 
Rabi oscillation of a logical qubit)
{\em in a fault-tolerant manner} \cite{NC00,Go97}. 
Although a non-fault-tolerant circuit can be much simpler,
such a circuit is too fragile to errors.
It is therefore important to explore new methods,
which are physically more feasible and natural, 
for inducing the encoded Hamiltonian evolution 
and thereby making QEC applicable.

In this paper, 
we propose such a new method,
choosing the Rabi oscillation as the  
Hamiltonian evolution to be protected. 
The method 
utilizes effective interactions that arise naturally among 
physical qubits.
We show that 
the values of the parameters in the interactions 
are to a great extent arbitrary.
Furthermore, 
switching of the interactions 
is unnecessary.
Therefore, a system of a logical qubit
with such interactions 
can be prepared 
easily by placing several two-level systems close to each other.
Once such a system is prepared, 
it is driven spontaneously and flawlessly by the Schr\"odinger equation.
This is much easier than to drive the system by a 
fault-tolerant quantum circuit.
On the other hand, 
we argue 
physically that it is highly probable that 
unwanted interactions should also exist in such a system.
While some of them are shown to be irrelevant, 
the others invalidate straightforward application of QEC.
As a resolution we present a protocol, 
which we call the error-correction sequence.
One can realize the protected Rabi oscillation 
by using the natural interactions (to induce the logical Rabi oscillation)
and a quantum circuit for the error-correction sequence.
This is much easier than realizing it wholly with a quantum circuit, 
because 
a fault-tolerant quantum circuit for inducing the logical Rabi oscillation,
which is the largest and most complicated part of the full circuit,
is unnecessary.

\section{Natural Hamiltonian for logical Rabi oscillation}
\label{sec:Hamiltonian}

We employ a two-level system 
as a basic element, 
which we call a qubit or {\em physical} qubit.
We represent operators acting on a qubit
in terms of the Pauli operators $X,Y,Z$ 
(i.e., $\sigma_1, \sigma_2, \sigma_3$), 
which are not necessarily those for a physical spin.
To apply QEC to the Rabi oscillation,
\begin{equation}
e^{i \omega X t} |0 \rangle
=
\cos(\omega t) |0 \rangle + i \sin(\omega t) |1 \rangle,
\label{eq:Rabi}\end{equation}
we replace a single qubit 
with a {\em logical} qubit which is composed of several qubits.
The basis states $|0 \rangle, |1 \rangle$
($+1$ and $-1$ eigenstate of $Z$, respectively)
of a qubit correspond to 
$|0_L \rangle, |1_L \rangle$ of a logical qubit.
The subspace (of the logical qubit) 
that is spanned by the latter 
is called the code space. 
For the reasons that will be described in Sec.\ref{sec:DandC}, 
we here take the Shor code \cite{Sh95}, in which 
a logical qubit is composed of nine qubits and
\begin{eqnarray}
|0_L\rangle &=&
{\bigl(|000\rangle+|111\rangle\bigr)
\bigl(|000\rangle+|111\rangle\bigr)
\bigl(|000\rangle+|111\rangle\bigr)
\over 2^{3/2}}, 
\\
|1_L\rangle &=& 
{\bigl(|000\rangle-|111\rangle\bigr)
\bigl(|000\rangle-|111\rangle\bigr)
\bigl(|000\rangle-|111\rangle\bigr)
\over 2^{3/2}}. 
\end{eqnarray}

We have to induce 
the {\em logical} Rabi oscillation; 
\begin{equation}
e^{i \omega X_L t} |0_L \rangle
=
\cos(\omega t) |0_L \rangle + i \sin(\omega t) |1_L \rangle, 
\end{equation}
where $X_L$ is a logical Pauli operator; 
$X_L |0_L \rangle = |1_L \rangle$ and 
$X_L |1_L \rangle = |0_L \rangle$.
Obviously, it can be induced 
if the Hamiltonian is $- \omega X_L$.
[Here and after, we take $\hbar=1$.]
This is an interaction among three or more qubits,
for any code that can correct all single-qubit errors 
(Appendix \ref{app:3 or more}).
For the Shor code, $X_L$ can be represented in various ways, 
e.g., as
$X_L = Z_3 Z_6 Z_9$ or 
$\Pi_{i=1}^9 Z_i$, 
where $Z_i$ acts on qubit $i$.
In the following, we take 
\begin{equation}
X_L = Z_1 Z_4 Z_7.
\end{equation}

\begin{figure}[tbp]
\begin{center}
 \includegraphics[width=0.4\linewidth]{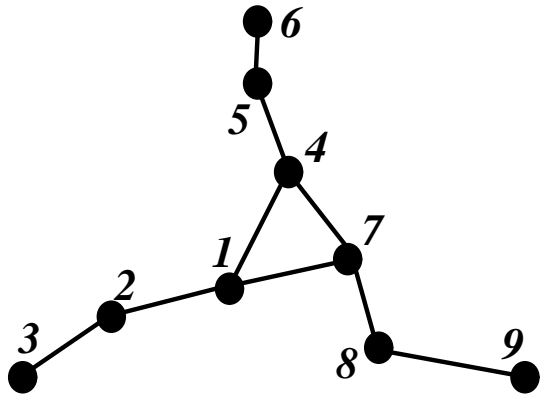}
 \caption{An example of the configuration of qubits that have 
the proposed Hamiltonian. 
The distances between the qubits are to a large extent arbitrary.
We label the qubits inside and outside the central triangle by
$r$ ($=1,4,7$) and $s$ ($=2,3,5,6,8,9$), respectively.
}
 \label{qubits configuration}
\end{center}
\end{figure}

Suppose that 
nine qubits
(such as atoms, quantum dots, and so on)
composing a logical qubit
are placed close to each other 
as shown in Fig.~\ref{qubits configuration}.
Then, as will be discussed in Sec.~\ref{sec:effectiveH},
a three-qubit interaction proportional to 
$X_L$ ($=Z_1 Z_4 Z_7$)
would be 
generated as an effective interaction.
[Similar three-qubit interactions were also discussed in
Refs.~\cite{3body1,3body2}.]
Unfortunately, however, if this interaction  
is strong enough 
unwanted two-qubit interactions
proportional to 
$Z_1Z_4, Z_4Z_7, Z_7Z_1$
should also be strong,
because otherwise the following unphysical conclusion 
would be drawn;
if one of qubits $1,4,7$ is removed
the other two qubits would have no interactions.
Furthermore, interactions between other pairs of qubits,
such as $Z_1 Z_2, Z_2 Z_3, \cdots$,
would also exist in general.
Therefore, 
a natural and simple Hamiltonian 
for the system of Fig.~\ref{qubits configuration} is
\begin{equation}
H = H_{D}+H_{S},
\label{total hamiltonian}
\end{equation}
where 
\begin{eqnarray}
&& \hspace{-3mm} H_{D} = -\omega Z_1 Z_4 Z_7 
-J(k_1 Z_1Z_4+k_4 Z_4Z_7+ k_7 Z_7Z_1),
\label{Rabi hamiltonian}\\
&& \hspace{-3mm} H_{S} = 
\sum_{s=2,3,5,6,8,9} g_s Z_{s-1} Z_s.
\label{Stabilizer hamiltonian}
\end{eqnarray}
Here, 
$\omega$, $J$, $k_r$'s, $g_s$'s
are real parameters. 
Since the signs of these parameters are irrelevant 
to the following discussions, 
we assume without loss of generality that they are positive.
Furthermore, since three-qubit interactions are generally 
weaker than two-qubit interactions
(see Sec.~\ref{sec:effectiveH}), 
we assume naturally that 
\begin{equation}
0< \omega \ll J.
\label{omega<<J}\end{equation}
Although single-qubit terms may also exist, 
we can forget them because, as discussed in Appendix \ref{Single terms}, 
they are irrelevant to the following discussions.

Note that the operators $Z_{s-1} Z_s$ 
in $H_{S}$ do not change $|0_L \rangle$ or $|1_L \rangle$, 
i.e., they are elements of the stabilizer \cite{Go97}
of the Shor code. 
Using this fact, 
we will show later 
by explicit calculations that
the values of $g_s$'s are irrelevant.
On the other hand, 
the two-qubit interactions 
in $H_D$ are not elements of the stabilizer,
and hence drive the state out of the code space.
Nevertheless, 
we will show in Sec.~\ref{sec:kk'} that 
the values of $J$ and $k_r$'s are fairly arbitrary as long as $\omega \ll J$.
The value of $\omega$ is also unimportant because
changing $\omega$ is just equivalent to changing the time scale.
Therefore, the values (including signs) of 
all the parameters in $H$ 
(hence the distances between the qubits)
are to a great extent arbitrary.
This makes our scheme robust to fabrication errors.
Once the system is thus fabricated, the law of the Nature
drives it flawlessly if noise is absent. 

\section{Difficulties and resolutions}
\label{sec:concepts}

We now discuss effects of noise.
There are two difficulties in applying 
QEC straightforwardly to the system driven by $H$.
We now explain them and 
resolutions.
For simplicity, we explain the case where 
$k_1=k_4=k_7=1$. 
More general cases will be discussed in Sec.~\ref{sec:kk'}.

\begin{figure}[t]
 \begin{center}
  \includegraphics[width=0.8\linewidth]{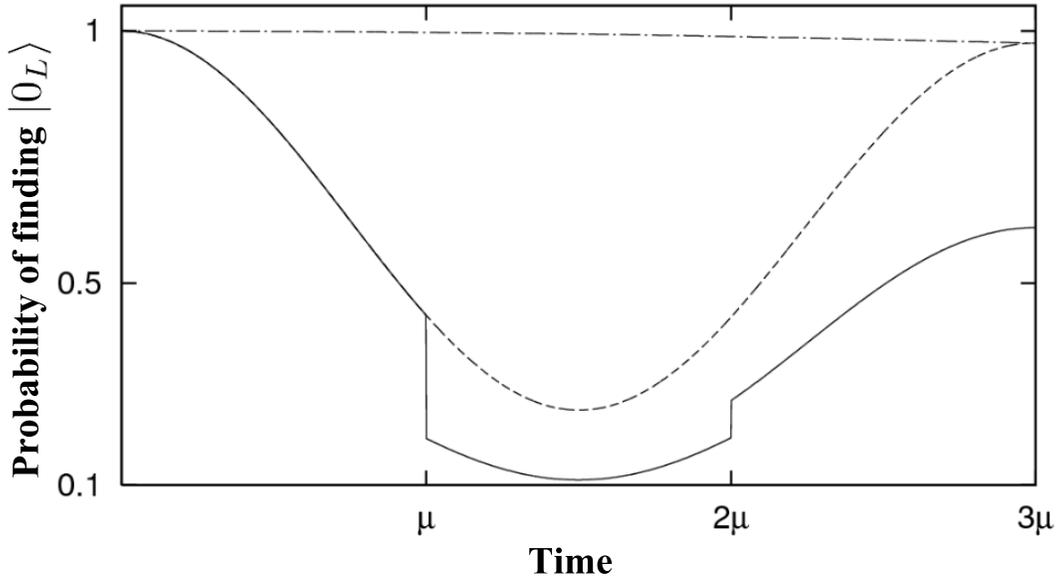}
  \caption{Probability of finding $|0_L\rangle$ plotted against time, 
  for the logical Rabi oscillation (chain line), 
  the Hamiltonian evolution by $H$ (dashed line), 
  that affected by QEC for phase errors (solid line), 
which is performed repeatedly with some intervals $\mu$. 
  }
\label{Phase synd meas mono}
 \end{center}
\end{figure}

We study the first difficulty by  
investigating the time evolution in the absence of noise,
i.e., we calculate 
$
|\psi(t) \rangle \equiv 
e^{-i H t}|\psi_L^0 \rangle
$,
where $|\psi_L^0\rangle$ is a vector in the code space.
We note that all terms in $H$ commute with each other, 
and that 
$H_S$ does not change 
$|\psi_L^0\rangle$
because all terms in $H_{S}$ are elements of the stabilizer. 
Using these facts and the relations 
$Z_1Z_4=Z_7X_L,Z_4Z_7=Z_1X_L,Z_7Z_1=Z_4 X_L$,
we find
\begin{align}
|\psi(t) \rangle
= &
\left\{\cos^3(Jt)-i\sin^3(Jt)\right\}
 e^{i\omega tX_L}|\psi_L^0 \rangle 
\nonumber\\  
& +\frac{i}{2}\sum_{r=1,4,7}e^{iJt}\sin(2Jt)
Z_r
X_L e^{i\omega tX_L}|\psi_L^0 \rangle.
\label{general period dynamics}
\end{align}
When $\sin(2J t) \neq 0$, 
this state is out of the code space
because of the last term.
Therefore, we cannot perform QEC for phase errors
at an arbitrary time,
because the syndrome measurement \cite{NC00}
to identify the errors
misidentifies the last term as a wrong term
generated by a phase-flip noise; 
if QEC for phase errors were performed with some intervals $\mu$ 
the time evolution 
would be affected as shown in Fig.~\ref{Phase synd meas mono}, 
even when noise is absent. 

\begin{figure}[t]
 \begin{center}
 \includegraphics[width=0.8\linewidth]{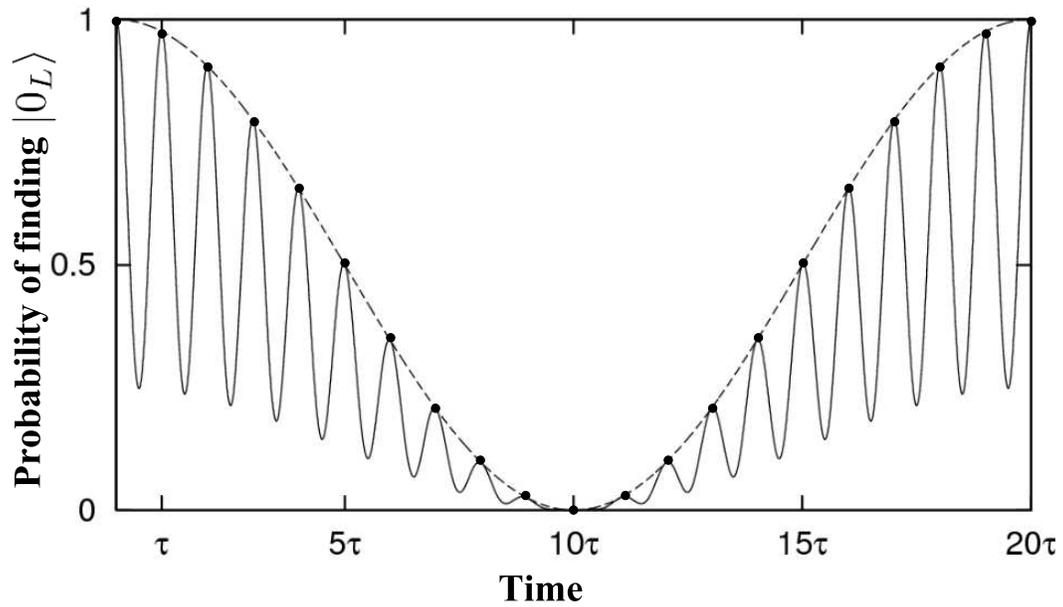}
\caption{Probability of finding $|0_L\rangle$ plotted against time,
for the logical Rabi oscillation (chain line)
and the Hamiltonian evolution by $H$ (solid line).
For clarity, we take $\omega /J$ ($\ll 1$) 
not so small; $\omega /J=0.1$. 
The dots represent the discrete logical Rabi oscillation.
}
\label{Discrete Rabi oscillation}
\end{center}
\end{figure}

However, 
if we focus on the discrete times
\begin{equation}
t_m \equiv m \tau 
\quad (m=0,1,2,\cdots),
\end{equation} 
then
$|\psi(t_m)\rangle$ is in the code space,
where 
\begin{equation}
\tau \equiv \pi/2J. 
\label{def:tau}\end{equation}
Therefore, 
we can perform QEC at $t=t_m$, 
for both phase and bit-flip errors. 
Furthermore, 
since
\begin{equation}
|\psi(t_m)\rangle=e^{i\omega t_m X_L}|\psi_L^0\rangle
= \left[ \cos(\omega t_m) + i \sin(\omega t_m)X_L \right]
|\psi_L^0\rangle
\end{equation}
apart from an irrelevant phase factor, 
the logical Rabi oscillation is realized 
at these discrete times, 
which we call the {\it discrete logical Rabi oscillation}. 
Since $\omega/J\ll 1$,
the intervals $\tau$  of the discrete times
are much shorter than the period $2 \pi/\omega$ of the 
Rabi oscillation.
Hence, 
the discrete logical Rabi oscillation
$\{|\psi(t_m)\rangle \}_{m=0,1,2\cdots}$ is quasi continuous
as shown by the dots in Fig.~\ref{Discrete Rabi oscillation}. 

To discuss the second difficulty,
let us study 
the time evolution in the presence of noise.
Suppose, e.g., that the system has evolved freely from noise
for $t<t'$, where $t_{m-1}<t'<t_m$,
until a bit-flip noise $X_1$ acts on qubit $1$ at $t'$.
Then the state at $t_m$ is evaluated as
\begin{eqnarray}
&& e^{-iH(t_m- t' )}X_1e^{-iH t' }|\psi_L^0\rangle  
\nonumber\\
&&  \quad
= e^{-i\left[g_2(2t'-t_m)+ \sum_{s \neq 2} g_st_m\right]}
X_1e^{iJ(2 t' -t_m)(Z_1Z_4+Z_7Z_1)}
\nonumber\\
&& \quad \quad \times (iZ_4Z_7)^m
e^{i \omega (2 t' -t_m)X_L}
 |\psi_L^0\rangle.
\label{state.X1}\end{eqnarray}
The terms proportional to $g_s$'s are irrelevant because
they contribute only to an overall phase factor.
Therefore, $g_s$'s may take arbitrary values.
The problem is that the above state is different 
from the correctable state $X_1|\psi(t_m)\rangle$, 
not only in the term generated by $Z_1Z_4+Z_7Z_1$ and $(iZ_4Z_7)^m$
but also in the wrong phase of the 
oscillation $\omega(2t'-t_m)$. 
That is, extra errors occur because the bit-flip error 
in qubit $1$ (or $4$ or $7$)
is `propagated' by $H$ to other qubits
\footnote{
Unlike the bit-flip errors, 
phase errors are not propagated by $H$. 
For example, 
if a phase error $Z_1$ acts on qubit $1$ at $t'$, 
the state at $t_m$ is 
$
e^{-iH(t_m- t' )}Z_1e^{-iH t' }|\psi_L^0\rangle
=Z_1e^{i\omega t_mX_L}|\psi_L^0\rangle
$ 
because $Z_1$ commutes with $H$. 
This state 
will be corrected by QEC for phase errors, which is performed at $t_m$. 
}. 
As a result, QEC at $t_m$ cannot recover the correct state.

To overcome this difficulty, 
we note that the syndrome measurement for 
bit-flip errors (unlike that for phase errors)
does not misidentify 
the state of Eq.~(\ref{general period dynamics})
as a wrong state. 
Hence, one can successfully 
perform QEC for bit-flip errors 
frequently (i.e., 
with intervals $\nu$ which are much shorter than $\tau$)
in the interval between $t_{m-1}$ and $t_{m}$ for all $m$.
As will be confirmed in the next section, 
this reduces the probability of errors small enough.

Our prescription is summarized as follows:
Perform QEC for both phase and bit-flip errors 
at all $t_m$'s (i.e., with intervals $\tau$), and 
perform QEC for bit-flip errors repeatedly 
with intervals $\nu$ ($\ll \tau$).
The latter intervals $\nu$ are not required to be regular.
We call this protocol the {\it error-correction sequence}.

\section{Effects of the error-correction sequence}
\label{sec:calculations}

To see how well the error-correction sequence 
protects the discrete logical Rabi oscillation 
against noise,
let us calculate the time evolution 
for $t_0 < t \leq t_1$, i.e., for $0 < t \leq \tau$,
quantitatively.

We divide the interval $(0,t]$ into $N$ subintervals; 
$(0,\Delta t], (\Delta t, 2\Delta t], (2\Delta t, 3\Delta t], \cdots$,
where $\Delta t \equiv t/N$.
We model noise by the product of 
depolarizing channels \cite{NC00} $\Pi_{i=1}^9 {\cal E}_{\Delta t}^{(i)}$,
where ${\cal E}_{\Delta t}^{(i)}$ 
acts on qubit $i$ at the end of every subinterval as 
\begin{equation}
{\cal E}_{\Delta t}^{(i)}[ \rho ]\equiv 
(1-\epsilon \Delta t )\rho+
{\epsilon \Delta t \over 3}
\sum_{\alpha=1}^3\sigma_{\alpha}^{(i)} \rho \sigma_{\alpha}^{(i)}.
\end{equation}
Here, 
$\rho$ denotes an input state, 
and
$\epsilon$ is a small positive parameter 
representing the strength of the interaction with the environment.
The initial state at $t=0$ is denoted by $\rho_L^0$, 
which is assumed to be in the code space.
We study its time evolution
up to the first orders in $\epsilon \tau$ and $\omega \tau$,
assuming that 
\begin{equation}
\epsilon \tau \ll 1
\mbox{ and } \omega \tau \ll 1,
\label{small_parameters}\end{equation}
where the latter comes from condition (\ref{omega<<J}).

If noise and QEC were absent, $\rho_L^0$ would evolve into
\begin{equation}
\rho_H(t) \equiv e^{-iH t} \rho_L^0 e^{iH t}
=e^{-iH_D t} \rho_L^0 e^{iH_D t}.
\end{equation}
When noise is present but QEC is not performed, on the other hand, 
$\Pi_{i=1}^9 {\cal E}_{\Delta t}^{(i)}$
acts at the end of every subinterval.
When $N=2$, for example, $\rho_L^0$ evolves into 
\begin{align}
 &
 \prod_{i=1}^9{\cal E}_{\Delta t}^{(i)}
 \left[e^{-iH\Delta t}
 \prod_{i=1}^9{\cal E}_{\Delta t}^{(i)}
 \left[\rho_H(\Delta t)\right]
 e^{iH\Delta t}
 \right]
 \\
 & =(1-9\epsilon t)\rho_H(t)
 +\frac{\epsilon}{3}\sum_{j=1}^2\Delta te^{-iH(t-j\Delta t)}
 \sum_{i=1}^9\sum_{\alpha=1}^3
 \sigma_{\alpha}^{(i)}\rho_H(j\Delta t)\sigma_{\alpha}^{(i)}
 e^{iH(t-j\Delta t)}. 
\end{align}
By taking $N\rightarrow \infty$, 
we obtain the state at $t$ without QEC as
\begin{align}
\rho(t,\rho_L^0)
\simeq & (1-9\epsilon t)\rho_H(t)
\nonumber\\ 
& +\frac{\epsilon}{3}
 \int_0^{t}dt'\, e^{-iH(t-t')}
 \sum_{i=1}^9\sum_{\alpha=1}^3
 \sigma_{\alpha}^{(i)}\rho_H(t') \sigma_{\alpha}^{(i)}
 e^{iH(t-t')}.
\end{align}
We calculate how this state is corrected by 
the error-correction sequence, in which 
bit-flip errors are corrected 
with intervals $\nu$ 
and both bit-flip and phase errors are corrected at $t=\tau$.
Although the intervals $\nu$ are not required to be regular, and 
\begin{equation}
n \equiv \tau/\nu
\label{def:n}\end{equation}
is not required to take an integral value, 
we here assume for simplicity that
$\nu$ is regular and $n$ is an integer.
We label qubits in and outside 
the central triangle of Fig.~\ref{qubits configuration}
by $r, r'$ ($=1,4,7$) and $s$ ($=2,3,5,6,8,9$), respectively.

At $t=\nu$, 
QEC for bit-flip errors is performed.
The pre-measurement state of the syndrome measurement
is $\rho(\nu,\rho_L^0)$.
The post-measurement state $\rho'(\nu)$
depends on the outcome of the syndrome measurement.
For example, when the bit-flip error in qubit $s$ is detected
(which happens with probability $2\epsilon \nu/3$),
\begin{equation}
\rho'(\nu) = 
\frac{1}{2} X_s \rho_H(\nu) X_s 
+\frac{1}{2} Y_s\rho_H(\nu) Y_s.
\end{equation}
By the recovery operation,
$\rho'(\nu)$ is changed into
\begin{equation}
\rho''(\nu) \equiv  X_s \rho'(\nu) X_s
=
\frac{1}{2} \rho_H(\nu)
+\frac{1}{2} Z_s\rho_H(\nu) Z_s, 
\end{equation}
which is a mixture of 
the correct state $\rho_H(\nu)$ and $Z_s\rho_H(\nu) Z_s$, 
the state with a phase error in qubit $s$.
At this stage, QEC for {\em phase} error is {\em not} performed because
$\rho_H(\nu)$ is out of the code space.

At $t=2 \nu$, 
QEC for bit-flip errors is performed again.
The pre-measurement state is $\rho(\nu, \rho''(\nu))$, 
where $\rho''(\nu)$ corresponds to one of possible outcomes of 
the previous syndrome measurement at $t=\nu$.
We can calculate $\rho'(2\nu)$
and $\rho''(2\nu)$ 
in the same way as we have calculated 
$\rho'(\nu)$ and $\rho''(\nu)$.
By repeating the arguments $n$ times, 
we obtain 
the probabilities of bit-flip errors during $0 < t <\tau$
and the corresponding states $\rho''(\tau)$ 
that are obtained at $t=n \nu = \tau$ by correcting the bit-flip errors.
To the first orders in $\epsilon \tau$ and $\omega \tau$, 
they are given by
\begin{equation}
\begin{array}{ccc}
  \hline
\mbox{error} & \mbox{probability}  & \mbox{corrected state } \rho''(\tau)\\
 \hline
\mbox{none}
  & 1-6\epsilon \tau
  & (1-3\epsilon \tau)\rho_H(\tau)+\frac{\epsilon \tau}{3}\sum_i Z_i\rho_H(\tau)Z_i, \\
X_s
  & 2\epsilon \tau/3
  & \frac{1}{2}\rho_H(\tau)+\frac{1}{2}Z_s\rho_H(\tau)Z_s, \\
X_r
  & 2\epsilon \tau/3
  & \frac{1}{2}\rho_e^{(r)}(\tau)+\frac{1}{2}Z_r\rho_e^{(r)}(\tau)Z_r,
\vspace{1mm}  
\\
\hline
\end{array}
\label{eq:t=tau-bf}\end{equation}
where 
\begin{equation}
\rho_e^{(r)}(\tau)
\equiv 
\int_0^{\nu}
e^{2iJ Z_r X_L t'}
\rho_H(\tau-2t') 
e^{-2iJ Z_r X_L t'} 
\frac{dt'}{\nu}.
\end{equation}

Finally at $t=\tau$, phase errors 
in $\rho''(\tau)$ 
are detected and corrected.
We denote the state after this QEC by $\rho'''(\tau)$.
Since $\rho''(\tau)$ depends on which qubit has suffered from 
a bit-flip error for $0 < t < \tau$,
so does $\rho'''(\tau)$.
If a bit-flip error has occurred in no qubit or in qubit $s$, 
$\rho'''(\tau)$ agrees with the correct state $\rho_H(\tau)$.
If, on the other hand, 
a bit-flip error has occurred in qubit $r$
(with probability $2\epsilon \tau/3$, see above), 
the conditional probability of each outcome of
the syndrome measurement for phase errors and 
the corresponding $\rho'''(\tau)$ are given by \footnote{
The phase flip in $r=1$ ($4, 7$) is equivalent to 
the phase flip in $s=2$ or $3$ ($5$ or $6$, $8$ or $9$) for the Shor code. 
}
\begin{equation}
\begin{array}{ccc}
  \hline
\mbox{error} & \mbox{probability}  
& \mbox{corrected state } \rho'''(\tau)\\
  \hline
  \mbox{none or } Z_r
  & 
\frac{3}{8}+\frac{1}{8}{\rm sinc}\frac{\displaystyle 4\pi}{\displaystyle n}
  & {\displaystyle 
\frac{a_n^{+}\rho_H(\tau)+a_n^{-}X_L\rho_H(\tau)X_L}
     {a_n^{+}+a_n^{-}}
     },
\\
Z_{r'} \ (r' \neq r)
  & 
\frac{1}{8}-\frac{1}{8}{\rm sinc}\frac{\displaystyle 4\pi}{\displaystyle n}
  & 
  \frac{1}{2} \left[\rho_H(\tau)+X_L\rho_H(\tau)X_L \right].
\vspace{1mm}  
\\
\hline
\end{array}
\end{equation}
Here, 
${\rm sinc}\, x \equiv (\sin x) /x$,
$
a_n^{\pm}
\equiv 
\frac{3}{16}
+\frac{1}{16}{\rm sinc}{4 \pi \over n}
\pm \frac{1}{4}{\rm sinc}{2 \pi \over n}
$, 
and terms of $O(\epsilon \tau)$ and $O(\omega \tau)$ have been dropped because
the probability that a bit-flip error has occurred 
is already of $O(\epsilon \tau)$.
By averaging $\rho'''(\tau)$ over all possible branches,
we obtain the average 
state $\rho_{c}(\tau)$ under the error-correction sequence as
\begin{equation}
 \rho_{c}(\tau)=\rho_H(\tau)
  -\epsilon \tau \left[ 1-{\rm sinc}\frac{2 \pi}{n} \right]
  \left[ \rho_H(\tau)-X_L\rho_H(\tau)X_L \right]. 
\end{equation}
Therefore, $\rho_{c}(\tau)$ approaches the correct state $\rho_H(\tau)$
with increasing $n$.
This can be seen more clearly from 
their trace distance \cite{NC00}, which is calculated for $n \gg 1$ as
\begin{equation}
{1 \over 2} {\rm tr} \left| \rho_c(\tau) -\rho_H(\tau) \right|
\simeq 2 \pi^2 \epsilon \tau L_{yz}(\tau)/3 n^2.
\end{equation}
Here, $L_{yz}(\tau)$ denotes 
the length of the 
projection onto the $y$-$z$ plane 
of the Bloch vector of $\rho_H(\tau)$
in the code space.
Hence, by taking 
\begin{equation}
n \gtrsim (1/\sqrt{\epsilon \tau}) \min \{ 1, \epsilon/\omega  \},
\label{eq:n}\end{equation}
we can reduce the distance to 
about 
$
6 L_{yz}(\tau) \max\{ (\epsilon \tau)^2, (\omega \tau)^2 \}
$.
Since $L_{yz}(\tau) = O(1)$, 
this is of the same order as the largest term that 
has been dropped in the above calculations.
That is, 
we have successfully recovered the correct state
at $t=\tau$ ($=t_1$), i.e., 
$\rho_{c}(t_1)=\rho_H(t_1)+
O\left(\max\{ (\epsilon \tau)^2, (\omega \tau)^2 \} \right)$.

In a similar manner, we can evaluate $\rho_{c}(t_m)$ by 
taking $\rho_{c}(t_{m-1})$ as the initial state,
and find that 
\begin{equation}
\rho_{c}(t_m)=\rho_H(t_m)+
O\left(\max\{ (\epsilon \tau)^2, (\omega \tau)^2 \} \right)
\end{equation}
for all $m$. 
Therefore, 
the discrete logical Rabi oscillation is protected, 
with only $O\left(\max\{ (\epsilon \tau)^2, (\omega \tau)^2 \} \right)$ 
probability of failure,
if we take $n$ as Eq.~(\ref{eq:n}). 
For example, we should take $n \gtrsim 10^2$ when 
$\epsilon \tau = \omega \tau =10^{-4}$.

\begin{figure}[tbp]
\begin{center}
\includegraphics[width=0.93\linewidth]{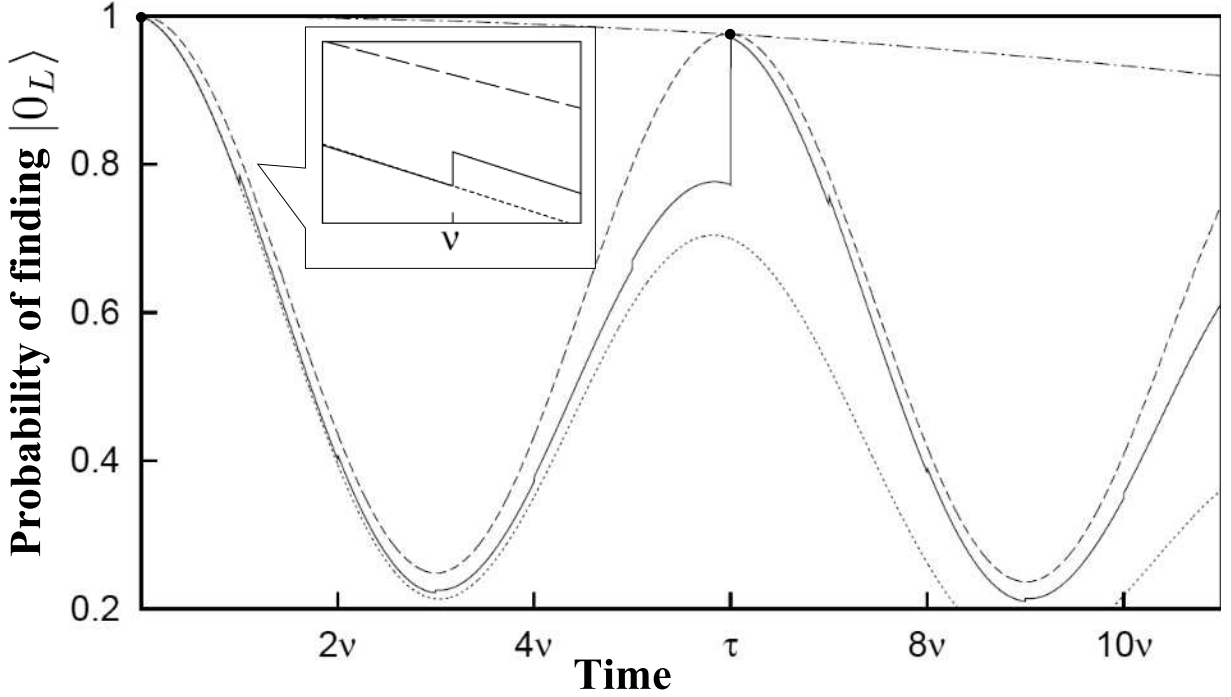}\end{center}
\caption{
Probability of finding $|0_L\rangle$ plotted against time
when $\rho_L^0=|0_L \rangle \langle 0_L|$,
for the logical Rabi oscillation (chain line), 
the Hamiltonian evolution by $H$ (dashed line), 
that affected by noise (dotted line), 
and that corrected by the error-correction sequence (solid line). 
The dots represent the discrete logical Rabi oscillation.
Here, $\omega/J=0.1$, $\epsilon \tau=\pi/100$, $\tau/\nu=6$.
Inset: magnification around $t=\nu$.
}
\label{Error-correction sequence}
\end{figure}

Figure \ref{Error-correction sequence} demonstrates 
how the error-correction sequence corrects errors, 
i.e., how the solid line approaches the dashed line.

\section{Arbitrariness of the parameters in $H_D$}
\label{sec:kk'}

It is clear from the results of 
Secs.~\ref{sec:concepts} and \ref{sec:calculations} 
that the value of $J$ is arbitrary as long as $\omega \ll J$.
On the other hand, we have assumed in those sections 
that $k_1=k_4=k_7=1$. 
In this section, 
we show that 
the error-correction sequence is successful also 
when $k_r$'s take other values. 

Recall that 
the error-correction sequence consists of two parts; 
QEC for both phase and bit-flip errors at all $t_m$'s, 
and 
QEC only for bit-flip errors with intervals $\nu$. 
The latter part is successful 
even when $k_r$'s are arbitrary real numbers,
because in general a Hamiltonian which does not contain 
$X_i$'s and $Y_i$'s, such as the proposed $H$, 
cannot flip the bit of any physical qubit.
Hence, the syndrome measurement for bit-flip errors 
does not misidentify the state evolved by such a Hamiltonian 
as a wrong state.
%

Regarding the former part, 
we start with showing that $k_r$'s can be arbitrary integers.
Note that 
QEC at $t_m$'s works well provided that 
the state of the qubits at $t_m$ would be in the code space 
if noise were absent. 
As discussed in Sec.~\ref{sec:concepts}, 
this condition is satisfied when $k_1=k_4=k_7=1$, 
because
$
|\psi(t_m)\rangle 
=e^{i\omega t_mX_L}|\psi_L^0\rangle 
$,
which is certainly 
in the code space.
When $k_r$'s are odd integers, we obtain the same result; 
\begin{eqnarray}
|\psi(t_m)\rangle
&=&
e^{-iH_Dt_m}|\psi_L^0\rangle
\nonumber\\
&=&
e^{i\omega t_mX_L}
e^{i\frac{m\pi}{2}(k_1 Z_1Z_4+k_4 Z_4Z_7+k_7 Z_7Z_1)}
|\psi_L^0\rangle
\nonumber\\
&=&
e^{i\omega t_mX_L}
\prod_{r=1,4,7}
\left[\cos\frac{m k_r \pi}{2}+i\sin\frac{m k_r \pi}{2} Z_r Z_{r+3} \right]
|\psi_L^0\rangle
\nonumber\\
&=&
e^{i\omega t_mX_L}|\psi_L^0\rangle, 
\end{eqnarray}
apart from irrelevant phase factors.
Here, $Z_{10} \equiv Z_1$, and we have used $(Z_1Z_4)(Z_4Z_7)(Z_7Z_1)=1$.
When $k_r$'s are general integers (not necessarily odd),
on the other hand, 
we have to add a certain procedure to the error correction sequence.
We explain this for the case where $k_1=1$ and either one of $k_4, k_7$ is even.
In this case, 
we find that 
\begin{equation}
|\psi(t_m)\rangle
=
\begin{cases}
e^{i\omega t_mX_L}|\psi_L^0\rangle
& \mbox{for even $m$}, 
\\
Z_rZ_{r'}e^{i\omega t_mX_L}|\psi_L^0\rangle
& \mbox{for odd $m$}.
\end{cases}
\end{equation}
Here, $r$ and $r'$ $(\neq r)$ each is
$1,4$ or $7$ depending on $k_4,k_7$.
For example, when $k_4$ is even and $k_7$ is odd,
$
|\psi(t_m)\rangle 
=Z_7Z_1e^{i\omega t_mX_L}|\psi_L^0\rangle
$ for odd $m$.
Although this state is out of the code space,
we note that the evolution into this state is 
not a stochastic process (such as evolution by noise)
but a deterministic process induced by the known Hamiltonian $H$ \footnote{
The values of all the parameters in $H$ 
can be measured experimentally after the system is fabricated.
}.
Hence, 
we can surely change this state to $e^{i\omega t_mX_L}|\psi_L^0\rangle$
by applying $Z_7Z_1$ just before QEC at $t_m$.
By adding this procedure to the error correction sequence,
we can successfully perform QEC at $t_m$'s.
Thus, 
the error-correction sequence,
supplemented with this additional procedure, 
works well when $k_r$'s are arbitrary integers.

Note that if $k_r$'s have a common factor $K$, one can redefine $k_r$'s and 
$J$ as
\begin{equation}
J' = KJ, \quad k'_r = k_r/K.
\end{equation}
The corresponding terms in $H_D$ are then rewritten as
\begin{equation}
J \sum_{r=1,4,7} k_r Z_r Z_{r+3}
=
J' \sum_{r=1,4,7} k'_r Z_r Z_{r+3}.
\end{equation}
Hence, one can
use $J'$ instead of $J$, 
which means, e.g., that $\tau' \equiv \pi / 2 J'$ is used 
instead of $\tau$.
The error correction sequence has such flexibility.

We next consider a more general case where $k_r$'s are rational numbers.
Suppose, for example, that $k_1=1, k_4=3/2, k_7=5/3$.
Then, one can redefine $k_r$'s and $J$ as
$J' = J/6, k'_r = 6k_r$,
and the corresponding terms in $H_D$ are rewritten as
\begin{equation}
J \sum_{r=1,4,7} k_r Z_r Z_{r+3}
=
J'(6 Z_1Z_4+9 Z_4Z_7+ 10 Z_7Z_1).
\end{equation}
Therefore, if one uses $\tau' \equiv \pi / 2 J'$ instead of $\tau$,
the error correction sequence is successful.
In general, 
if there exists a real number $\kappa $ such that 
$\kappa  k_r$'s are integers and 
\begin{equation}
J' \equiv J/\kappa \gg \omega,
\label{eq:J/kappa}\end{equation}
then the error correction sequence is successful
if one uses $\tau' \equiv \pi / 2 J'$ instead of $\tau$.

Finally, we consider the case where 
$k_r$'s are irrational numbers.
We note that an irrational number can be well approximated by  
rational numbers.
When $k_1 = \pi$ ($=3.14159\cdots$), for example, it can be approximated by 
$22/7$ ($=3.14285\cdots$), 
$333/106$ ($=3.14150\cdots$), 
and so on.
Let $k_{1*}$ be such a rational number.
The difference $k_1 - k_{1*}$ is negligible if 
$J \left| k_1 - k_{1*} \right| t \ll 1$.
Therefore, for the time interval $t$ that satisfies
\begin{equation}
t \ll 1/J \left| k_1 - k_{1*} \right|,
\label{upperlimit-t}\end{equation}
this case reduces to the one where 
$k_r$'s are rational numbers.
If one takes $k_{1*}$ such that 
$\left| k_1 - k_{1*} \right|$ is smaller,
the upper limit of $t$ given by Eq.~(\ref{upperlimit-t})
becomes longer, whereas condition (\ref{eq:J/kappa})
becomes harder to satisfy because the denominator of $k_{1*}$ becomes
greater.

To summarize this section, 
the error-correction sequence 
works well for fairly arbitrary values of $k_r$'s.
Although it is better that one can successfully fabricate 
the system in such a way that $k_r$'s are integers, 
one can also accept most  
systems which have non-integral values 
of $k_r$'s (because of fabrication errors).
This fact makes the preparation of the system easier.

\section{Derivation of the effective interactions}
\label{sec:effectiveH}

The proposed Hamiltonian $H$ consists of  
Ising-type interactions and three-qubit interactions 
among physical qubits. 
We here discuss 
how they are generated as effective interactions 
from more fundamental interactions. 

Many physical systems can be candidates for physical qubits 
that have the proposed $H$.
As an example, 
we here 
consider quantum dots in a semiconductor \cite{BDEJ95,BW98}. 

To be more concrete, we assume that the spin of an electron in a dot is 
polarized by a high external magnetic field, 
so that we can forget about the spin degrees of freedom.
We also assume that 
the potential barrier between the dots is high and thick
so that electron tunneling between the dots is negligible.
This and (possibly) the Coulomb interaction, by which 
states with two electrons in a single dot have 
much higher energies than states with a single electron,
exclude double occupancy of a dot.
For single-electron states of a dot, 
we assume that  
only the ground and the first excited states, 
denoted by $|0\rangle$ and $|1\rangle$, are relevant 
because higher states 
have much higher energies and/or the transition matrix elements 
to them are small.
As a result, we can treat each dot as a system with two quantum levels,
$|0\rangle$ and $|1\rangle$, i.e., as a qubit.
For the reasons that will be explained below, 
we also assume that 
all dots in a logical qubit are asymmetric and 
different (in size and/or shape) 
so that accidental degeneracy is lifted.

The effective Hamiltonian $H_{\rm eff}$ for a set of 
such qubits (dots) 
is the sum of single-qubit terms and the effective interactions.
The effective interactions 
are derived from more elementary interactions 
$V, W, \cdots$, 
which are {\em effective} interactions among {\em conduction} electrons
in {\em homogeneous} bulk semiconductors.
On the other hand, 
$V, W, \cdots$ are derived from 
even more elementary interactions, such as the Coulomb interactions
between electrons in vacuum. 
Since two- and three-body interactions have been studies in many physical
systems (see, e.g., Refs.~\cite{SCH03,BMZ07}),
we here consider a two-body interaction $V$
and a three-body interaction $W$.
Generally, the latter is much weaker than the former
%
\footnote{
For example, 
when the dielectric constant $\epsilon$ for 
the screened Coulomb interaction 
$U=e^2/\epsilon|{\bm r}-{\bm r}'|$
between electrons, which are located at ${\bm r}$ and ${\bm r}'$,
weakly depends on the location ${\bm r}''$ of another electron
as $\epsilon=\epsilon_0+\delta \epsilon({\bm r}'')$
($\epsilon_0 \gg |\delta \epsilon({\bm r}'')|$), 
then $U \simeq V+W$, where 
$V \equiv e^2/\epsilon_0|{\bm r}-{\bm r}'|$ and
$W \equiv -e^2 \delta \epsilon({\bm r}'') /(\epsilon_0)^2|{\bm r}-{\bm r}'|$.
This $W$ is much weaker than $V$ because 
$\left| W/V \right| = \left| \delta \epsilon/\epsilon_0 \right| \ll 1$.
}.
Since four- or more-body interactions are even weaker, 
we neglect them.

We can represent $H_{\rm eff}$ as a polynomial 
of the Pauli operators.
In general, it 
would have terms that include 
$X_i\equiv 
|0\rangle_i \, \null_i \langle 1 |+| 1 \rangle_i \, \null_i \langle 0 |$
and $Y_i\equiv 
-i| 0 \rangle_i \, \null_i \langle 1 |+i| 1 \rangle_i \, \null_i \langle 0 |$,
where the subscript $i$ ($=1, 2, \cdots$) labels the qubits.
Such terms are non-diagonal terms 
that are proportional to 
$|n\rangle \langle m|$ $(m\neq n)$,
where $|n\rangle$ and $|m\rangle$ 
are product states of $| 0 \rangle_i $'s and $| 1 \rangle_i $'s
(such as $\prod_i | 0 \rangle_i $).
As discussed in Refs.~\cite{BB04} and \cite{VC04} and in 
Appendix \ref{sec:XY}, 
contributions from the non-diagonal terms 
to the time evolution are negligible if
\begin{equation}
\left| \frac{\langle n |(V+W)|m\rangle}{\Delta E_{nm}} \right| \ll 1
\quad \mbox{for every $n,m$ ($\neq n$)},
\label{eq:xi=small}\end{equation}
where $\Delta E_{nm}$ is the difference 
in energy of single qubit terms 
between $|n\rangle$ and $|m\rangle$. 
[A more precise expression of this condition is given 
in Appendix \ref{sec:XY}, where 
$\langle n |H'|m\rangle$ corresponds to 
$\langle n |(V+W)|m\rangle$.]

In typical situations,
$V$ and $W$ are significant only between
{\em adjacent} dots
(such as dots $1, 4, 7$, dots $1, 2$, dots $2, 3$, and so on,
of Fig.~\ref{qubits configuration})
because $V$ and $W$ generally decrease as the distance is
increased.
In such a case, 
one can make $|\Delta E_{nm}|$
larger than $\left| \langle n |(V+W)|m\rangle \right|$ 
by making the sizes and/or shapes of adjacent dots different.
One can also make $|\Delta E_{nm}|$ larger
by modulating spatially the magnitude of the external magnetic field.
If condition (\ref{eq:xi=small}) is satisfied by these methods, 
one can drop non-diagonal terms, and hence $H_{\rm eff}$
reduces to $H$, which consists only of
$Z_i=| 0 \rangle_i \, \null_i \langle 0 |-| 1 \rangle_i \, \null_i \langle 1 |$'s,
when considering the time evolution.

On the conditions and assumptions mentioned above, 
$H$ can be derived simply by taking the diagonal matrix elements, 
between $| n \rangle$'s,
of the effective Hamiltonian for conduction electrons,
\begin{equation}
H_0^{\rm el}+V+W,
\label{H3:org}\end{equation}
where $H_0^{\rm el}$ denotes the non-interacting part, 
which includes the confining potential of the dots.
We here present explicit results
for the three qubits in 
the central triangle of Fig.~\ref{qubits configuration}.
Interactions between the other qubits 
can be derived more easily in a similar manner.
%

Since the potential barrier is high, 
the wavefunctions $\psi_r^0({\bm r})$ and $\psi_r^1({\bm r})$ 
of $|0\rangle_r$ and $|1\rangle_r$, respectively, 
are well localized within each dot. 
As a result, 
overlap of the wavefunctions of different dots is negligibly small, 
i.e., $\psi_r^a({\bm r})\psi_{r'}^b({\bm r})\simeq 0$ for $r\neq r'$ 
and for all $a,b$ ($=0,1$). 
Using this fact, we find that 
the effective Hamiltonian is given by
\begin{equation}
-\frac{1}{2}\sum_{r=1,4,7}\zeta_r Z_r-\sum_{r>r'}J_{rr'}Z_rZ_{r'}-\omega Z_1Z_4Z_7,
\label{H3:eff}\end{equation}
where, for $a,b,c=0,1$,
\begin{gather}
\begin{split}
\zeta_1=\zeta^0_1
-\frac{1}{2}
\sum_{a,b} (-1)^{a}(V_{a \bullet b}+V_{a b \bullet})
-\frac{1}{4}\sum_{a,b,c}(-1)^aW_{abc}
\label{eq:zeta}
\end{split} 
\\
\begin{split}
 J_{14}=-\frac{1}{4}\sum_{a,b}(-1)^{a+b}V_{ab\bullet}
 -\frac{1}{8}\sum_{a,b,c}(-1)^{a+b}W_{abc},
\end{split}
\\
\omega = -\frac{1}{8}\sum_{a,b,c}(-1)^{a+b+c}W_{abc},
\end{gather}
and similarly for the other $\zeta_r$'s and $J_{rr'}$'s.
Here, 
$\zeta_r^0$ is the energy difference between 
$|1 \rangle_r$ and $|0 \rangle_r$,
and
\begin{eqnarray}
V_{ab\bullet} &\equiv& 
\iint \left|\psi_{1}^{a}({\bm r}) \right|^2
V({\bm r},{\bm r}')
\left| \psi_{4}^{b}({\bm r}') \right|^2
d{\bm r}d{\bm r}',
\\
W_{abc} &\equiv&
\iiint 
W({\bm r},{\bm r}',{\bm r}'')
\left| \psi_{1}^{a}({\bm r}) \right|^2
\left| \psi_4^{b}({\bm r}') \right|^2
\left| \psi_7^{c}({\bm r}'') \right|^2
d{\bm r}d{\bm r}'d{\bm r}'',
\end{eqnarray}
and similarly for $V_{a \bullet b}, V_{\bullet ab}$.
In fact, one can easily verify that all the diagonal matrix elements 
of Eq.(\ref{H3:org}), between $| n \rangle$'s, 
agree with those of Eq.(\ref{H3:eff}).

It is seen that 
the single-dot energy $\zeta_r$ is renormalized by 
the interactions $V$ and $W$,
and the two-qubit effective interactions are generated from $V$ and $W$, 
whereas the three-qubit effective interaction is generated from $W$. 
Regarding the magnitudes of the effective coupling constants, 
$\omega$ is much smaller than $J_{rr'}$'s 
because the former is derived only from the weaker interaction $W$.
Note that 
$\omega$ does not vanish by accidental degeneracy
because we have assumed that 
all dots in a logical qubit are asymmetric and different.

Since we can forget about the single-qubit terms (i.e., the first term 
of Eq.~(\ref{H3:eff})) as discussed in 
Appendix \ref{Single terms}, 
Eq.~(\ref{H3:eff}) agrees with the proposed $H_D$, 
Eq.~(\ref{Rabi hamiltonian}), 
where $J_{rr'}$ correspond to $k_r J$.

\section{Discussions and Conclusions}
\label{sec:DandC}

We have shown in Secs.~\ref{sec:concepts} and \ref{sec:calculations}
that two-qubit interactions in 
$H_D$ cause errors which are correctable not by 
the straightforward application of QEC but by 
the error-correction sequence.
One might expect that such errors could be corrected more easily 
by using
more elaborate codes such as the one in Ref.~\cite{Ruskai}.
If such codes are used, however, 
$X_L$ in $H_D$ becomes 
an interaction among 
three or more qubits.
Generally, 
if $l$ qubits are crowded to induce an $l$-qubit interaction
corresponding to $X_L$,
unwanted interactions among $l'$ ($<l$) qubits 
are also induced,
as we have discussed on $H$.
For any code that can correct all single-qubit errors,
some of such unwanted interactions are {\em not} elements of the stabilizer 
\footnote{
For example, suppose that 
$X_L=X_1X_4X_7$ 
and unwanted interactions are $X_1X_4$, $X_4X_7$, and $X_7X_1$. 
If these unwanted interactions were elements of the stabilizer, 
we would have 
$|1_L\rangle = X_L|0_L\rangle =X_1 X_4 X_7 |0_L\rangle =X_1 |0_L\rangle$,
which shows that $X_1$ would be another expression of $X_L$.
However, this is impossible 
for any code that can correct all single-qubit errors 
because, as discussed in Appendix \ref{app:3 or more}, 
$X_L$ should be a three- or more-fold tensor product of the Pauli operators.
}.
If $l' \geq 3$ like the code of Ref.~\cite{Ruskai}, 
they cause errors which cannot 
be corrected even by the error-correction sequence. 
If $l'<3$ like the Shor code and the Steane code \cite{St96}, 
they can be dealt with the error-correction sequence. 

We have also shown that the values of $g_s$'s in $H_S$ are arbitrary.
Such 
great flexibility would not be obtained if 
we employed a non-degenerate code \cite{NC00},  
because its stabilizer
does not include two-fold tensor products of the Pauli operators. 
For example, 
the Steane code 
is a non-degenerate code 
and hence it has less flexibility. 
For these reasons, 
we have employed in this paper the Shor code, 
which is a degenerate code with 
$l=3$ (because we can take $X_L=Z_1 Z_4 Z_7$) and $l'=2$.

Possibility of use of other codes is worth exploring.
It is also worth exploring the possibility of 
replacing a circuit for the syndrome measurements 
with 
another natural interactions.
Our preliminary study indicates that this is basically possible,
and more detailed studies are in progress.
Furthermore, it is interesting to apply the present idea to general 
time evolutions (such as general SU(2) rotations)
and/or to general systems 
(such as systems composed of many logical qubits).
A possible way of realizing this may be mixed use of 
an Hamiltonian (such as the one of this paper) and
simple quantum circuits.
This might also be applicable to quantum simulations 
\cite{Feynman, Lloyd}.
Since these subjects are beyond the scope of the present paper, 
we leave them as subjects of future studies.

In conclusion, 
we have shown that 
the Rabi oscillation of a logical qubit encoded by the Shor code
can be induced by 
a Hamiltonian that is composed of 
natural short-range interactions among physical 
qubits (Sec.~\ref{sec:Hamiltonian}).
The Hamiltonian replaces 
the most complicated part of a quantum circuit that 
would be necessary for inducing and protecting the logical Rabi oscillation.
More specifically, 
the state driven by the proposed Hamiltonian 
agrees with the logical Rabi oscillation at discrete times 
$t_m=m\tau$ ($m=0,1,2,\cdots$),
which is quasi continuous as shown in Fig.~\ref{Discrete Rabi oscillation}. 
We call it the discrete logical Rabi oscillation (Sec.~\ref{sec:concepts}).
To prepare a physical system that has the proposed Hamiltonian,
one has simply to place two-level systems (which are used as physical qubits), 
such as asymmetric quantum dots (Sec.~\ref{sec:effectiveH}), 
as shown in Fig.~\ref{qubits configuration}. 
The parameters of this system, 
such as the positions and the sizes of the dots,
are to a great extent arbitrary
because the proposed Hamiltonian has great flexibility
(Secs.~\ref{sec:Hamiltonian} and \ref{sec:kk'}).
This makes the fabrication of the system easier.
Once the fabrication is finished, 
one can measure the coupling constants of the effective interactions, 
and the important parameters such as $\tau$ can be calculated from them. 
To protect the discrete logical Rabi oscillation against noise, 
the ordinary QEC cannot be applied straightforwardly.
However, 
we have shown that it can be protected 
by a new protocol, which we call the error-correction sequence 
(Secs.~\ref{sec:concepts} and \ref{sec:calculations}).
In this protocol, 
QEC for both phase and bit-flip errors is performed at $t_m$'s, 
whereas QEC only for bit-flip errors is performed frequently 
in the interval between $t_{m-1}$ and $t_m$ for all $m$. 
The frequency of the latter is determined by 
the strength of noise and the parameters of the effective 
interactions (Sec.~\ref{sec:calculations}). 
One can realize the protected Rabi oscillation 
by using the natural Hamiltonian (to induce the logical Rabi oscillation)
and a quantum circuit for the error-correction sequence.
This is much easier than realizing it 
wholly with a fault-tolerant quantum circuit.

\begin{acknowledgments}
The authors thank Y. Matsuzaki for discussions.
This work is partly supported by KAKENHI.
\end{acknowledgments}

\appendix 

\section{$X_L$ is an interaction among three or more qubits}
\label{app:3 or more}

Let $P_c$ be the projection operator onto the code space;
\begin{equation}
P_c =|0_L\rangle \langle 0_L|+|1_L\rangle \langle 1_L|.
\end{equation}
An $n$-qubit code which 
can correct all single-qubit errors 
satisfies the following condition \cite{NC00};
\begin{equation}
P_c \sigma_{\alpha}^{(i)} \sigma_{\beta}^{(j)} P_c
=\chi_{i \alpha, j \beta} P_c
\quad (i,j=1,\cdots , n; \ \alpha,\beta=0,1,2,3).
\label{eq:Pc}\end{equation}
Here, 
$\sigma_{\alpha}^{(i)}$ denotes the identity ($\alpha=0$)
and Pauli ($\alpha=1,2,3$) operators acting on qubit $i$,
and 
$\chi_{i \alpha, j \beta}$ is an element of 
some Hermitian matrix.

If $X_L$ were a Pauli operator or 
a two-fold tensor product of Pauli operators, 
Eq.~(\ref{eq:Pc}) could not be satisfied. 
For example, if $X_L=X_1 X_2$ for some code 
the left-hand side of Eq.~(\ref{eq:Pc}) with 
$\sigma_{\alpha}^{(i)}=X_1, \sigma_{\beta}^{(j)}=X_2$ 
(i.e., $i=1, j=2, \alpha=\beta=1$)
reduces to 
\begin{equation}
P_cX_1X_2P_c=P_cX_LP_c
=|0_L\rangle \langle 1_L|+|1_L\rangle \langle 0_L|.
\end{equation}
Since this is neither vanishing nor proportional to 
$P_c$, 
Eq.~(\ref{eq:Pc}) is not satisfied for any value of 
$\chi_{11}^{12}$. 
This means that 
such a code cannot correct all single-qubit errors. 

Therefore, 
$X_L$ is a three- or more-fold tensor product of the Pauli operators
(which corresponds to an interaction among three or more qubits)
for any code that can correct all single-qubit errors.

\section{Irrelevance of single-qubit terms}
\label{Single terms}

When two levels of a qubit have different energies,
a single-qubit term, which represents the energy difference, 
arises in its effective Hamiltonian as discussed in 
Sec.~\ref{sec:effectiveH}.
All effects of such single-qubit terms can be canceled 
if we do everything in the rotating frame \cite{VC04}.
Although this fact seems to be known widely, 
we here explain it in order to 
clarify its meaning in 
the context of QEC. 

Let us investigate the time evolution of a state $\rho_+$ 
by the following Hamiltonian 
\begin{equation}
H_+
=-{1 \over 2}\sum_{i=1}^9\zeta_i Z_i + H
=-{1 \over 2}\sum_{i=1}^9\zeta_i Z_i + H_D+H_S, 
\end{equation}
where $\zeta_i$'s are real numbers. 
We can go to the rotating frame (an interaction picture) 
by $U_0\equiv \exp({i \over 2}\sum_{i=1}^9 \zeta_iZ_i t)$, as
$
\rho^{\text{rot}}=U_0^{\dagger} \rho_+U_0
$.
It evolves according to 
\begin{equation}
i\frac{d}{dt}\rho^{\text{rot}}=[U_0^{\dagger} H U_0, \rho^{\text{rot}}]=[H,\rho^{\text{rot}}], 
\end{equation}
where we have used $[H,U_0]=0$. 
Therefore, $\rho^{\text{rot}}$ undergoes the same unitary evolution 
as that of $\rho$ of Sec.~\ref{sec:calculations}. 
Furthermore, it is easy to show that the depolarizing channel in the rotating frame is 
also the same as the one in Sec.~\ref{sec:calculations}. 
Thus, 
in the presence of noise, 
$\rho^{\text{rot}}$ evolves in the same manner as 
$\rho$ of Sec.~\ref{sec:calculations}. 
Therefore, 
the error correction sequence
will be successful 
if we set the initial state $\rho^{\text{rot}}(0)$ in the code space 
and perform QEC in the rotating frame.

For example, 
the observables for the syndrome measurement
in the rotating frame 
are 
$
M_{b_1}^{\text{rot}} \equiv Z_1Z_2
$,
$
M_{p_1}^{\text{rot}} \equiv X_1X_2X_3X_4X_5X_6
$,
and so on.
In the laboratory frame (Schr\"odinger picture), 
they are given by 
$
M_{b_1}=U_0 Z_1Z_2 U_0^{\dagger}=Z_1Z_2
$ 
and 
$
M_{p_1}=U_0 X_1X_2X_3X_4X_5X_6 U_0^{\dagger}
=\prod_{i=1}^6
\exp({i \over 2}\zeta_i Z_i t)X_i\exp(-{i \over 2}\zeta_i Z_i t)
$,
respectively.

\section{Irrelevance of terms including $X, Y$}
\label{sec:XY}

It seems widely accepted by researchers of NMR
that the non-diagonal terms,
which include $X_i$'s and/or $Y_i$'s,
in $H_{\rm eff}$ 
are irrelevant to the time evolution 
if condition (\ref{eq:xi=small}) is satisfied
(see, e.g., Refs.~\cite{BB04} and \cite{VC04}).
For completeness, we here show that
this is indeed true 
under reasonable assumptions.

Let us decompose $H_{\rm eff}$ as
\begin{equation}
H_{\rm eff}
=
H_0 + H + H',
\quad
H_0 \equiv - {1 \over 2}\sum_i\zeta_i Z_i, 
\end{equation}
where 
$\zeta_i$ is the energy difference 
(that is renormalized, like Eq.~(\ref{eq:zeta}),
by interections among dots)
between $| 1 \rangle_i $ and $| 0 \rangle_i$, 
$H$ is a polynomial of $Z_i$'s only,
and $H'$ consists of the other terms
(such as $X_1 Y_1$, $X_1 X_2 Z_3$, and so on)
which include $X_i$'s and/or $Y_i$'s.

We denote a product state of $| 1 \rangle_i $'s and $| 0 \rangle_i $'s,
such as $\prod_i | 1 \rangle_i $,
by $|n \rangle$.
In terms of such product states,
$H_0$ and $H$ are diagonal,
whereas $H'$ gives the off-diagonal elements.
To characterize the magnitude of the latter, 
we define the parameter $\xi_{nm}$ by
\begin{equation}
\xi_{nm}
\equiv
\begin{cases}
\displaystyle
{\langle n | H' | m \rangle 
\over
\Delta E_{nm}
}
& \mbox{if } \langle n | H' | m \rangle \neq 0,
\\
0
& \mbox{if } \langle n | H'| m \rangle = 0,
\end{cases}
\end{equation}
where $\Delta E_{nm}$ denotes the difference
of the eigenvalues of $H_0$ between  
$|n\rangle$ and $|m\rangle$.
We also define
\begin{equation}
\bar{\xi} \equiv
\sqrt{\sum_{n,m} \left| \xi_{nm} \right|^2}.
\end{equation}

Consider the time evolution operator $U_{\rm eff}(t)$ 
generated by $H_{\rm eff}$.
We can write it as
\begin{equation}
U_{\rm eff}(t)
\equiv
\exp \left( -i H_{\rm eff} t \right)
= U(t) e^{-i Q(t)}, 
\end{equation}
where 
\begin{equation}
U(t) \equiv e^{- i (H_0+H) t}
= e^{- i H_0 t} e^{- i H t}, 
\end{equation}
and $Q(t)$ is the Hermitian operator defined by
$\displaystyle e^{-i Q(t)} \equiv U^\dagger(t) U_{\rm eff}(t)$.
It is clear that 
\begin{equation}
Q(t) = 0 \mbox{ when } \bar{\xi}=0.
\label{eq:Q=0}\end{equation}
If $\bar{\xi}$ were large then $Q(t)$ would be significant,
particularly when $\Delta E_{nm} = 0$ for all $n,m$,
for which $\bar{\xi}=\infty$.
On the other hand, if $\bar{\xi}$ is small enough
$\| Q(t) \|$ is expected to be small,
where $\| \cdot \|$ denotes the operator norm.
It is natural to assume that 
\begin{equation}
\mbox{Assumption 1: } 
\mbox{$Q(t)$ is continuous in $\xi_{nm}$'s
in the neighborhood of $\bar{\xi}=0$}.
\label{eq:Q=cont}
\end{equation}
This assumption seems reasonable from the 
perturbation expansion of the time evolution
operator in the interaction picture,
which corresponds to 
$e^{i H_0 t } U_{\rm eff}(t) = e^{- i H t} e^{-i Q(t)}$;
\begin{eqnarray}
&&
1 -i t \sum_{n}
\langle n| H |n\rangle |n\rangle \langle n|
 -i\sum_{n} \sum_{m \, (\neq n)}
\int_0^{t}e^{i\Delta E_{nm}t'}\,dt'
 \langle n | H' |m\rangle |n\rangle \langle m|
 +\cdots
\nonumber\\
&&=
1 -i t \sum_{n}
\langle n| H |n\rangle |n\rangle \langle n|
-\sum_{n} \sum_{m \, (\neq n)}
\left( e^{i\Delta E_{nm}t} - 1 \right)
\xi_{nm} |n\rangle \langle m|
 +\cdots,
\label{eq:expansion}\end{eqnarray}
each term of which is continuous with respect to $\xi_{nm}$.

Assumption 1, together with Eq.~(\ref{eq:Q=0}),
means that 
for any small positive number $\varepsilon$
there exists a positive number $\bar{\xi}_{\varepsilon, t}$ such that
\begin{equation}
\| Q(t) \| < \varepsilon
\mbox{ for all } \bar{\xi} < \bar{\xi}_{\varepsilon, t}.
\end{equation}
In other words, for a given time period $[0, t)$
we can neglect $Q(t)$,
i.e., we can regard $U_{\rm eff}(t)=U(t)$,
if $\bar{\xi}$ is small enough.
This means that the time evolution by $H_{\rm eff}$ 
takes place as if $|n\rangle$'s 
(which are eigenstates of $H_0+H$)
were its eigenstates.
That is, if we expand an initial state in terms of $|n\rangle$'s
as $\sum_{n} c_n |n\rangle$, 
\begin{equation}
e^{-i H_{\rm eff} t}
\sum_{n} c_n |n\rangle
\simeq
\sum_{n} c_n 
e^{-i \langle n|(H_0+H)|n\rangle t}
|n\rangle.
\end{equation}

Note that 
the above argument is general in the sense that we have not assumed 
any specific forms for $H$ and $H'$.
For example, the argument in Appendix A of Ref.~\cite{BB04},
where specific forms have been assumed,
is essentially a special case of the present general argument.

In the above argument, 
we have not excluded the possibility that 
$\bar{\xi}_{\varepsilon, t}$ increases with increasing $t$.
This will not cause difficulty when one sets an upper limit of $t$.
To be more complete, however, we here discuss 
dependence of $\bar{\xi}_{\varepsilon, t}$ on $t$.
We note that the coefficients of the third term of 
Eq.~(\ref{eq:expansion}) are upper bounded as
\begin{equation}
\left| \left( e^{i\Delta E_{nm}t} - 1 \right)\xi_{nm} \right|
\leq
2 \left| \xi_{nm} \right|
\leq
2 \bar{\xi}
\end{equation}
for all $t$.
This is due to the fact that 
$t$ appears only through the oscillatory factor $e^{i\Delta E_{nm}t}$.
Since this is the case also for higher-order terms,
we expect that 
\begin{equation}
\mbox{Assumption 2: } 
\mbox{$\bar{\xi}_{\varepsilon, t}$ has an upper bound $\bar{\xi}_\varepsilon$,
which is independent of $t$}.
\end{equation}
If this is true, then 
for any small positive number $\varepsilon$
and for all $t$
\begin{equation}
\| Q(t) \| < \varepsilon
\mbox{ for all } \bar{\xi} < 
\bar{\xi}_\varepsilon.
\label{Q<xie}\end{equation}
In other words,
we can regard $U_{\rm eff}(t)=U(t)$
even for long $t$
if $\bar{\xi}$ is small enough.

\end{document}